\documentclass[%
 reprint,
 amsmath,amssymb,
 aps,
superscriptaddress,prl,floatfix]{revtex4-2}

\usepackage{graphicx}
\usepackage{dcolumn}
\usepackage{bm}

\usepackage{xcolor}
\usepackage{verbatim}
\usepackage{caption}
\usepackage{lineno}

\begin{document}

\preprint{APS/123-QED}


\title{
Internal stress drives ferromagnetic-like ordering in networks of proliferating bacteria}

\author{Nicola Pellicciotta}
\affiliation{NANOTEC-CNR, Soft and Living Matter Laboratory, Piazzale A. Moro 5, Roma, 00185, Italy}
\affiliation{%
 Dipartimento di Fisica, Sapienza Università di Roma, Piazzale A. Moro 5, Roma, 00185, Italy}%
 
\author{Luca Angelani}
\affiliation{%
Istituto dei Sistemi Complessi - Consiglio Nazionale delle Ricerche (ISC-CNR), Piazzale A. Moro 2, I-00185 Roma, Italy}
\affiliation{%
 Dipartimento di Fisica, Sapienza Università di Roma, Piazzale A. Moro 5, Roma, 00185, Italy}%

\author{Roberto Di Leonardo}%
\affiliation{%
 Dipartimento di Fisica, Sapienza Università di Roma, Piazzale A. Moro 5, Roma, 00185, Italy}%
\affiliation{NANOTEC-CNR, Soft and Living Matter Laboratory, Piazzale A. Moro 5, Roma, 00185, Italy}

\date{\today}

\begin{abstract}
Proliferation is a defining feature of life. Through growth, division, and death, living systems consume energy and inject mass, breaking conservation laws and driving collective phenomena from biofilm formation to embryonic development. Yet, while active matter physics has advanced our understanding of self-propelled agents, quantitative frameworks for proliferating systems are still emerging, and most work focuses on simplified settings. Here, we study \textit{E.coli} bacteria growing inside a network of single-file microchannels as a minimal model of structured environments. Competition for free volume drives the spontaneous emergence of coherent growth patterns that persist across generations but vanish when the channel links exceed the typical cell size at birth. Despite the strongly out-of-equilibrium character of the dynamics, the observed phenomenology can be quantitatively captured by an effective equilibrium description in which the flow state at each node is represented by a spin variable with ferromagnetic interactions. Simulations of growing elastic cells show that this coupling arises from internal stress accumulated at network nodes due to dynamical constraints. Our results reveal a surprising correspondence between proliferating active matter and equilibrium statistical mechanics, highlighting open fundamental questions and offering a first step toward describing growth phenomena in real-world complex environments.
\end{abstract}

\maketitle


Proliferation is more than the biological act of making cell copies. It provides a physical mechanism for spatial exploration and for the emergence of form, from biofilm formation to morphogenesis \cite{d1917growth,nadell2016spatial,davies2023mechanisms}. 
When compared with motility, growth-driven dynamics are extremely slow, yet they can generate very large mechanical stresses~\cite{montel2012stress,delarue2016self,kroeger2011regulator}. While motile cells move by microns per seconds ~\cite{lisicki2019swimming,bechinger2016active} by generating thrust forces in the pN range~\cite{lauga2009hydrodynamics}, growing cell collectives expand with speeds that are typically hundred times smaller but with pressures reaching up to $10^5$ pN/$\mu$m$^2$~\cite{alric2022macromolecular,blanc2024bacterial}. From a more fundamental physics standpoint, if self-propelled systems may violate time-reversal symmetry in subtle ways \cite{o2022time,nardini2017entropy}, growing systems are dramatically irreversible as they locally inject mass and degrees of freedom into the system \cite{hallatschek2023proliferating}.
Active matter dynamics can thus arise from fundamentally different microscopic drives, leading to distinct collective behaviors.
Self-propelled particles tend to constantly move in a space that is obstructed by neighbors and external walls \cite{bechinger2016active}. This leads to local alignment \cite{zhang2010collective,vicsek2012collective}, polar order \cite{marchetti2013hydrodynamics}, and large-scale collective flows\cite{dombrowski2004self,perez2025bacteria,bricard2013emergence}. By contrast, proliferating cells actively inject mass through growth and division, and therefore they organize in structures that maximize local packing, while simultaneously allowing coordinated rearrangements to efficiently evacuate space and accommodate new cells \cite{bi2015density,mongera2018fluid,wittmann2023collective,dell2018growing,you2019mono,karita2022,cadart2019physics}. While active matter physics has revealed many of the fundamental principles governing velocity-driven organization in self-propelled collectives \cite{jorge2024active,solon2015pressure,cates2015motility}, the search for quantitative frameworks describing the density-driven dynamics of growing cells is less explored and largely hindered by the lack of simple experimental systems that could allow the formulation and testing of basic underlying rules \cite{hallatschek2023proliferating}.
The case of bacteria is particularly emblematic and of direct practical relevance, as they proliferate in a wide variety of spatially structured environments in nature, ranging from soils and riverbeds \cite{ebrahimi2014microbial,dang2016microbial} to biological tissues such as the lung \cite{kirch2012optical,mancini2025hyphal} or the gut \cite{donaldson2016gut}. These habitats are inherently heterogeneous and consist of interconnected structures spanning several orders of magnitude, down to the micron scale \cite{nadell2016spatial}, which profoundly influence their growth, size and division through mechanical and chemical feedback mechanisms \cite{blanc2024bacterial, dufrene2020mechanomicrobiology,scheidweiler2024spatial,taheri2015cell,mannik2009bacterial}. 
Beyond its theoretical relevance, building a general framework capable of describing bacterial proliferation under such realistic conditions would have significant implications for ecology \cite{raynaud2014spatial}, infection biology \cite{anderson2003intracellular,justice2004differentiation}, and the design of innovative drug-delivery strategies \cite{gurbatri2022engineering,forbes2010engineering}.\\
Here we study \textit{E.coli} bacteria growing inside networks of interconnected microchannels with variable lengths. We find that, as bacteria compete for access to exit channels, the system stochastically switches between metastable growth modes that allow continuous cell growth while minimizing accumulated stress. The discrete nature of growth introduces both a time and a length scale in the problem. 
While the division time sets the timescale for switching dynamics, when the length of network edges exceeds the cell size, correlations are progressively lost.
Although the system is strongly out of equilibrium, its statics and dynamics are well captured by an equilibrium model in which node states are represented by spin variables that interact via effective ferromagnetic couplings encoding accumulated elastic energy. Numerical simulations further support this framework, revealing a previously unexplored analogy between bacterial proliferation in confined networks and equilibrium statistical-mechanical systems.

\begin{figure*}[t]
\centering
\includegraphics[width=0.95\textwidth]{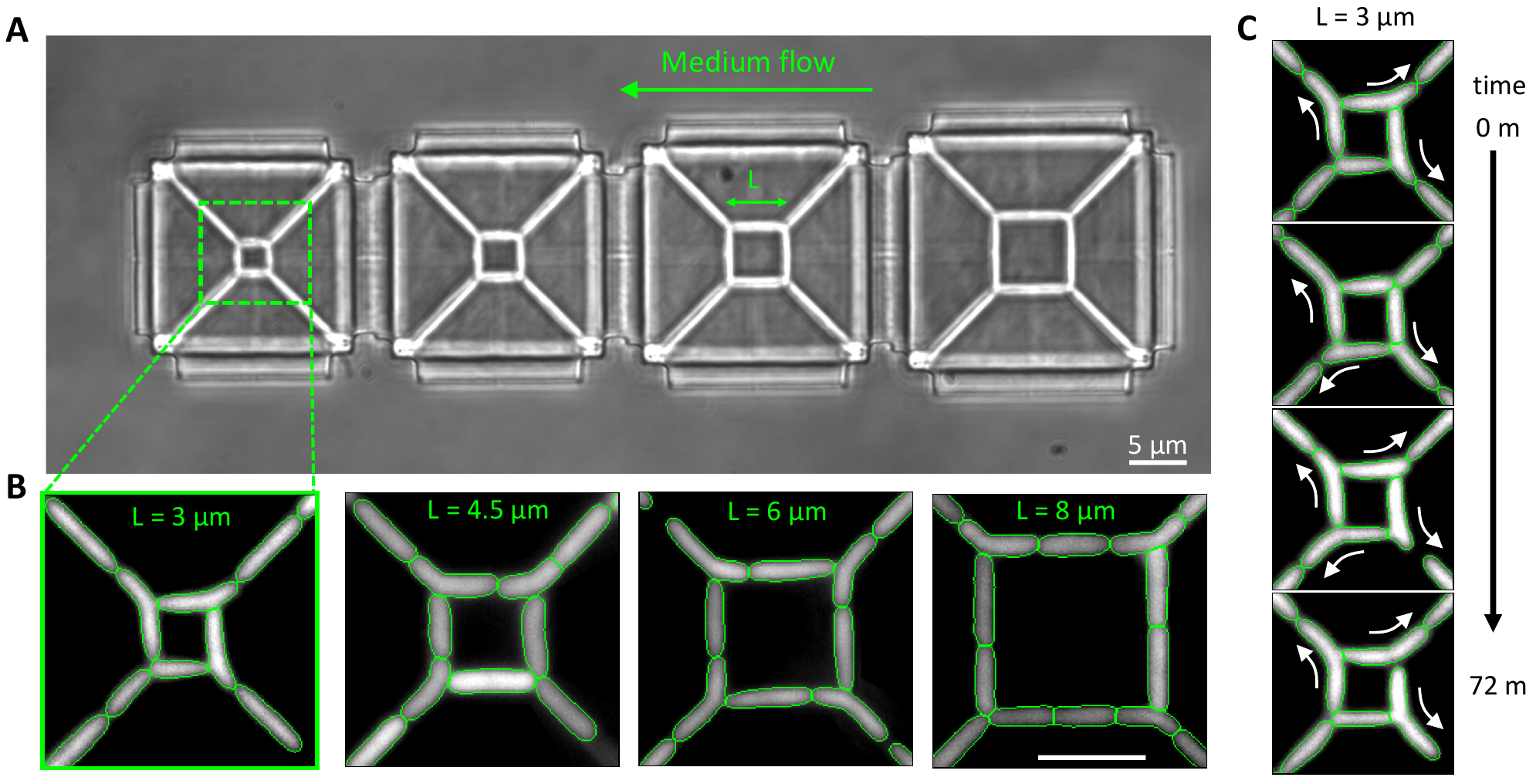}
\caption{ \textbf{Transition from ordered growth in networks of microchannels.} (A) Bright-field image of the SU-8 microchambers containing the square network features. (B) Fluorescence images of GFP-expressing bacteria growing in microchambers of different sizes. Scalebar 5$\mu$m.(C) Timelapse images showing ordered bacterial growth within a network of size L=3 $\mu$m, with arrows indicating the output flux of bacteria. The absence of an arrow indicates that the node is temporarily unoccupied. } 
\label{fig:fig1}
\end{figure*}

\subsection{Bacterial growth in microfabricated networks.} 
The “mother machine” is a microfluidic device consisting of independent parallel microchannels that constrain bacteria to grow in single file, and was originally developed for the quantitative study of cell size regulation in bacteria \cite{taheri2015cell}. Our microstructures can be thought of as interconnected networks of mother machines, designed to highlight collective effects in cells growing within a shared, complex environment where they compete for space.
As a minimal network structure we chose a simple 4-node cycle with pendant edges (Fig.~\ref{fig:fig1}A). Using two-photon lithography \cite{maruo2008recent,vizsnyiczai2017light}, we fabricated square networks of microchannels in SU-8 photoresist, with oblique exits (45$^\circ$) located at each vertex . The microchannels have cross-sectional dimensions of approximately $0.7~\mu\mathrm{m}$, enforcing single-file growth and passage.
Each network is connected to nanometric outlet channels that allow nutrient diffusion while preventing bacterial escape (see Extended Data Fig.~1), thereby maintaining sustained growth under geometric confinement. On the same substrate, we fabricated multiple networks with varying side length  $L = [3,\,4.5,\,6,\,8]~\mu\mathrm{m}$, while keeping the outlet channel length fixed ($L_{\mathrm{out}} = 10~\mu\mathrm{m}$). The patterned substrate was bonded to a PDMS chip incorporating a $2~\mathrm{mm}$-wide and $200~\mu\mathrm{m}$-high flow channel, allowing controlled bacterial loading and continuous nutrient perfusion. A dilute suspension of motile, GFP-expressing \textit{E. coli} was then introduced into the device, rapidly colonizing the microchannel networks, where cells remained stably confined and initiated proliferation. Bacterial growth was monitored by time-lapse fluorescence microscopy, with images acquired every 4 minutes using a $100\times$ objective (Fig.~\ref{fig:fig1}B). Within this hybrid microfluidic system, sustained nutrient perfusion supported continuous bacterial growth over several days.
Under these conditions, bacteria grew exponentially, with an average growth rate $\alpha \approx 0.6 \pm 0.2~\mathrm{h}^{-1}$ (corresponding to a doubling time $\tau_d \approx 1.1 \pm 0.3~\mathrm{h}$) and a characteristic cell length at birth of $\ell \approx 3.3 \pm 0.7~\mu\mathrm{m}$ (Extended Data Fig.~2). Further details on bacterial growth within the networks, image acquisition, and cell tracking are provided in the Methods.\\
Since bacteria continuously grow and locally inject mass in the network, cell flow will  through the outlet channels. Because the microchannels are only wide enough to accommodate single-file passage, strong competition arises at each network node, where adjacent cells compete for access to the available exits. By tracking bacterial growth inside the networks, we quantified the flux of cells exiting through each vertex of the square. Remarkably, we found that for short edge lengths, particularly $L = 3\,\mu\mathrm{m}$, bacteria exhibit a highly ordered pattern of escape (Fig.~\ref{fig:fig1}C). In this regime, cell growth and movement become coordinated, allowing each bacterium to exit the chamber without mutual interference. As the network size $L$ increases, however, this ordered behavior progressively breaks down (Supplementary Video 1). In larger networks, continuous cell division creates intermittent gaps between adjacent bacteria; when these gaps reach a vertex, they provide an opportunity for a neighboring cell to overtake the exit. As a consequence, some cells experience compression and are unable to grow outward, while others intermittently exploit available openings and expand through one or two of the exits.\\
%
\begin{figure*}[t!]
\centering
\includegraphics[width=0.9\textwidth]{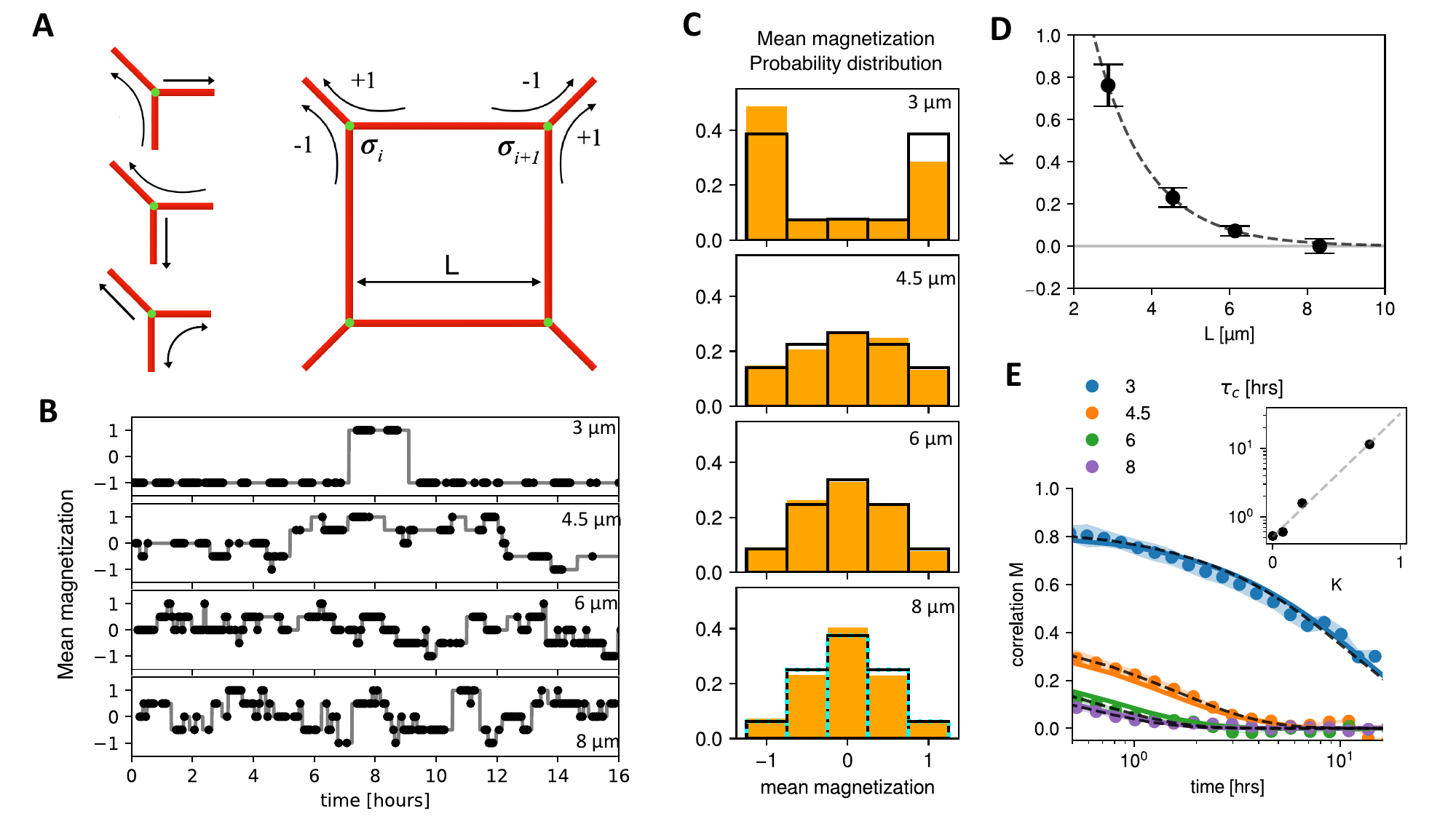}
\caption{ \textbf{Bacterial flux dynamics in microchannel networks: experiments and model.}(A) Left: Possible flows in a corner node. Right: Schematic of the spin model applied to the output flow of bacteria in the network. 
(B) Mean magnetization over time in networks of different sizes (single experiment). Black markers indicate the measured values, while the gray line represents an interpolation applied when no bacteria are passing through the network node.
(C) Probability distribution of the mean magnetization from experimental data (orange) and from the model (black lines). The cyan dotted line represent the binomial distribution for four statistically independent spin variables.  Data were collected from more than ten independent experiments for each network size \(L\).
(D) Coupling parameter \(K\) derived from the model as a function of the network length \(L\). The dashed line is an exponential fit to guide the eye.
(E) Time correlation function of the mean magnetization for the four network sizes, represented by round markers. The shaded area shows the standard error of the mean. Each data point is averaged over more than ten independent experiments. Dashed lines represent exponential fits, while solid lines show Monte Carlo dynamics extracted using the corresponding \(K\) values from the statics. The inset highlights the exponential growth of the characteristic time \(\tau_c\) with the parameter \(K\).
}
\label{fig:fig2}
\end{figure*}
%
\subsection{Mapping bacterial flow onto an equilibrium spin model.}
\begin{figure}[h]
\centering
\includegraphics[width=0.45\textwidth]{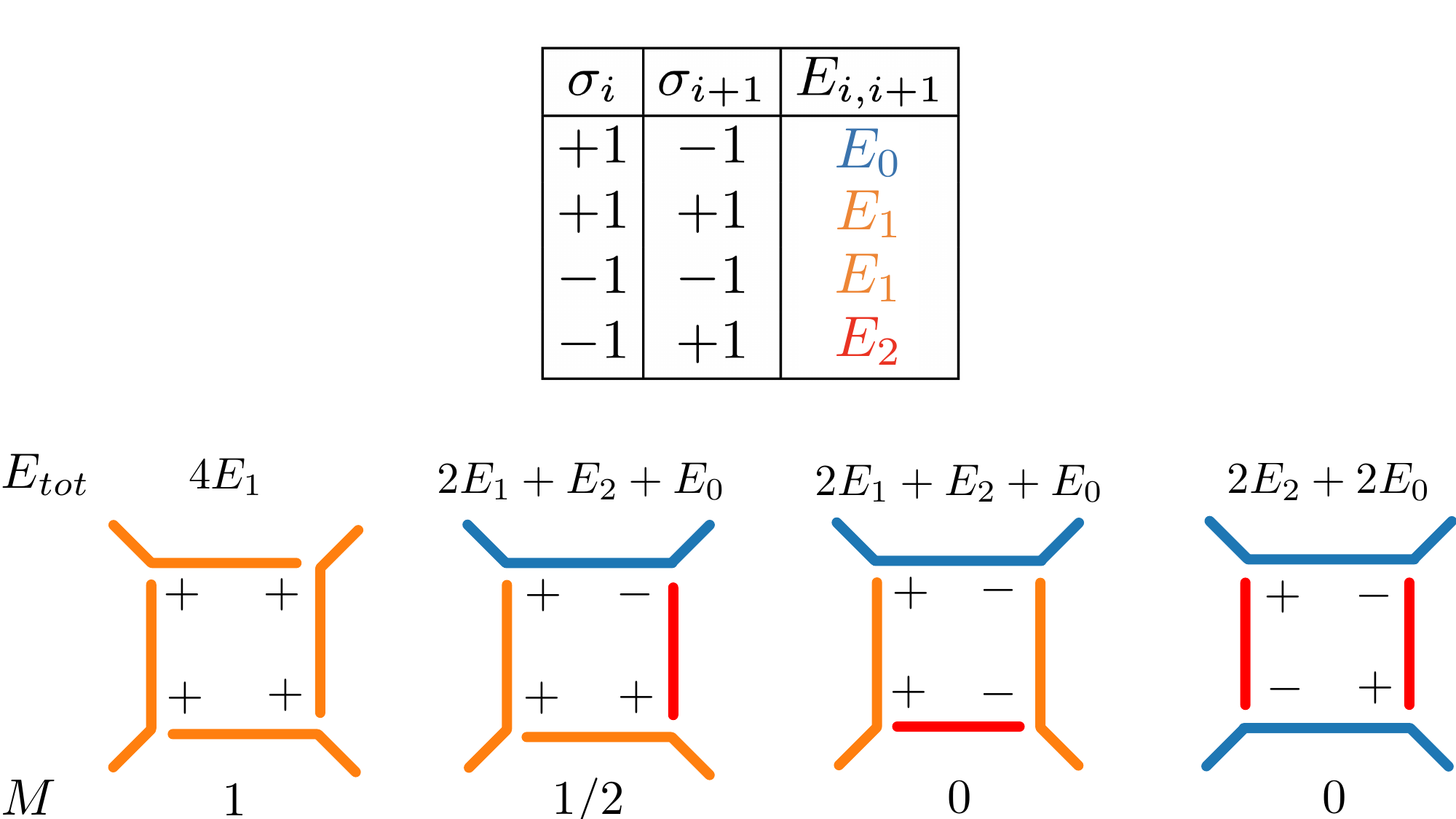}
\caption{ \textbf{Energy associated with spin configurations.} The table summarizes the interaction energy associated with each configuration of adjacent spins.  Illustrative examples of network configurations are shown in the bottom panel. The value shown above each schematic denotes the associated total energy, while the value below corresponds to the mean magnetization $M$. For symmetry reasons, we show only representative configurations with non-negative magnetization, as configurations with reversed spins are energetically equivalent.}
\label{fig:fig3}
\end{figure}

Each edge of the network acts as a source, injecting mass into the system at a rate solely constrained by mechanical compression. For an isolated edge bounded by two open-end nodes, this mass is expelled symmetrically at a rate proportional to the edge's total length. When multiple edges share the same node, the law of mass conservation states that the sum of the mass currents flowing into the node must be zero. This is the same Kirchhoff's law that applies to electric and fluidic networks, but there is a fundamental difference in the present case. When cells move in a single file, all mass is encapsulated by cell walls so currents cannot split at a node. Thus, solutions to Kirchhoff's law are restricted to cases in which all the mass injected into a node by one edge flows entirely into a second edge, while all the other edges contribute zero current. Consider, for example, a corner node in our network. Three edges share the node, so the restricted Kirchhoff's law only admits three possibilities, as shown in Fig.\ref{fig:fig2}A. The bottom case requires strong right-angle bending of the cell and is rarely observed. Consequently, only two states remain for every node: positive, where cells circulate clockwise, and negative, where they circulate counterclockwise.  
Any growth state in our network is therefore fully specified by four spin variables $\sigma_i = \pm 1$, each associated with one of the  four network nodes indexed by $i=1,2,3,4$ (Fig.~\ref{fig:fig2}A). We analyzed time lapse movies of cells growing in square networks of different size $L$ and extracted the time series of the four spin variables $\sigma_i(t)$. 
In Fig.2B we report the observed time evolution of the  mean ``magnetization" $M$:
\begin{equation}
M(t) = \frac{1}{4}\sum_{i=1}^4\sigma_i(t)    .
\end{equation}
Missing data points in the time series correspond to situations where the spin value is undefined, such as when one node is temporarily not occupied by a cell. For a four-node network $M$ can only assume the five distinct values $-1,-1/2,0,1/2,1$ with $\pm1$ corresponding to a highly symmetric flow configuration as shown, for example, in Fig.\ref{fig:fig1}C -- third panel (see also the diagram at the bottom left in Fig.\ref{fig:fig3}). 
For the smallest network length,  $M$ seems to remain trapped in the two metastable states $M=\pm 1$, with only occasional transistions between the two. As the network length increases, spin values start to decorrelate and $M$ fluctuates between all intermediate states more rapidly. To quantify and compare the degree of order of cell growth in networks of different sizes, we analyze the probability distributions of the magnetization $P(M)$, obtained by averaging over multiple independent experiments, Fig.~\ref{fig:fig2}C. For the smallest system size, $P(M)$ is bimodal, with pronounced peaks at $-1$ and $+1$, indicating that the system spends roughly 80\% of the observation time in these two states, with the remaining time uniformly distributed among intermediate values of $M$.
As the network size increases, the distribution gradually develops a single peak at $0$ until, for the largest network size, the shape of $P(M)$ approaches the binomial distribution expected for four statistically independent spin variables (dotted line in Fig.\ref{fig:fig2}C).
This phenomenology suggests the presence of a coupling mechanism that favors the alignment of adjacent spins, with a strength that decreases as the system size increases. To gain insight into the nature of this mechanism, we can focus on a single edge and analyze all possible spin configurations at its two end nodes. 
When both spins are positive or negative, the edge will have an available exit at one node while the other one will be blocked. When spins have opposite signs, two distinct configurations occur. For example, let's focus on the top edge in Fig.\ref{fig:fig2}A. If the left node is positive and the right node is negative, then new cells can be expelled at both ends, and the compression is partially relieved.  Otherwise, if the left node is negative and the right node is positive, both exits are blocked for the top edge, 
and a growth pressure will build up. We assign energies $E_0, E_1, E_2$ to each of this three states to quantify their progressive degree of internal stress.
If we number spins progressively as we move clockwise around the square, we can build a table associating each adjacent spin configurations to an energy level (Fig.\ref{fig:fig3}).
From this table we can deduce a generic expression for the interaction energy between spins $i$ and $i+1$:
\begin{align}
E_{i,i+1} &= \frac{E_2+2E_1+E_0}{4} 
- \frac{E_2-E_0}{4} (\sigma_i-\sigma_{i+1}) \notag \\
&\quad - \frac{E_2-2E_1+E_0}{4} \, \sigma_i \sigma_{i+1} .
\end{align}
When summing over the four nodes to obtain the total configuration energy $E_{tot}$, the second term cancels out,
 while the first contributes only an irrelevant global energy offset $E_{off}=E_2+2E_1+E_0$. 
Therefore, the total energy can be written as $E_{tot}=E_{off} +E$, with
\begin{equation}
    E = - J \sum_{i=1}^4 \sigma_i \sigma_{i+1} ,
    \label{eq:H}
\end{equation}
where 
periodic boundary conditions are considered, i.e. $\sigma_5\equiv \sigma_1$, and 
\begin{equation}
    J=\frac{E_2-2E_1+E_0}{4}.
\label{eq:K}
\end{equation}

The interaction energy, as well as the total energy of the system, can take only three discrete values, which are reported in Fig.~\ref{fig:fig3}, together with some associated representative configurations.
When $J>0$,  Eq. (\ref{eq:H}) is the Hamiltonian of a ferromagnetic Ising model. This condition holds provided that the total energy of the ordered state (all spins equal) is lower than that of the state with two blocked edges (alternating spins), i.e. $4E_1 < 2(E_2 + E_0)$.
The states $M = \pm 1$, which are the most probable in small structures, also correspond to the lowest energy level of the spin Hamiltonian. We now ask whether an effective equilibrium model can quantitatively reproduce the full shape of the observed probability distributions despite the fact that the system is strongly out of equilibrium.  
Assuming a Boltzmann-like statistics, with an effective inverse temperature parameter $\beta$, the partition function reads
\begin{equation}
    Z = \sum_{\{\sigma\}} e^{-\beta E} = 12 + 4 \cosh{4K} ,
\end{equation}
where $K = \beta J$.
Within this effective equilibrium description, the mean magnetization $M$ takes the values $\{-1,-1/2,0,1/2,1\}$ with probability  
\begin{equation}
P(0) = \frac{4+2e^{-4K}}{Z} , ~~~~
P(\pm1/2) = \frac{4}{Z} , ~~~~
P(\pm 1) = \frac{e^{4K}}{Z} ,
\label{eq:probs}
\end{equation}
resulting in the mean values  
\begin{equation}
\label{m2eq}
\langle M \rangle = 0\;,\;\;\langle M^2 \rangle = \frac{1+e^{4K}}{6+2\cosh 4K} .
\end{equation}
By inverting the second expression, we can determine the value of the coupling parameter $K$ corresponding to the experimentally measured values of $\langle M^2 \rangle$ for each system size $L$. This value of $K$ can then be substituted into Eq.~\ref{eq:probs} in order to compare the resulting equilibrium predictions with the observed  probability distributions of $M$. Fig.~\ref{fig:fig2}C demonstrates that our effective equilibrium model provides a quantitatively accurate description of at least the static properties of the system.
Overall, we found that the coupling parameter $K$ decreases exponentially with the network length $L$, approaching almost zero for the largest $L$ (Fig.~\ref{fig:fig2}D). 

\subsection{Equilibrium theory reproduces growth dynamics.} The dynamics of the network over time can be characterized  by the time-correlation function of the mean magnetization $C(t) = \langle M(t_0)M(t_0+t) \rangle_{t_0}$
(Fig.~\ref{fig:fig2}E). 

All measured correlations can be well matched with an exponential decay $A e^{-t/t_c}$ (dashed line), with a characteristic time $t_c$ 
that decreases from $11$ hours to around $30$ minutes as the network size grows from $3$ to $8$ $\mu$m. More interestingly, the characteristic time $t_c$ scales exponentially with the coupling parameter  $K=\beta J$ (insert), suggesting that equilibrium thermally activated processes could play a role in governing the system's dynamics.
To test this hypothesis we implemented a Monte Carlo (MC) dynamics using the Metropolis algorithm for the effective Hamiltonian in Eq.\ref{eq:H}.
Simulating spin dynamics for the coupling values $K$ extracted from experiments, we obtain predictions for the correlation functions $C(t)$ for all system sizes. Remarkably, these theoretical correlations accurately reproduce all experimental curves once we set the time scale for a Metropolis step to the global value $\tau_{_{MC}} = 0.4$ hours (Fig.~\ref{fig:fig2}E). 

\begin{figure*}[h!]
\centering
\includegraphics[width=0.8\textwidth]{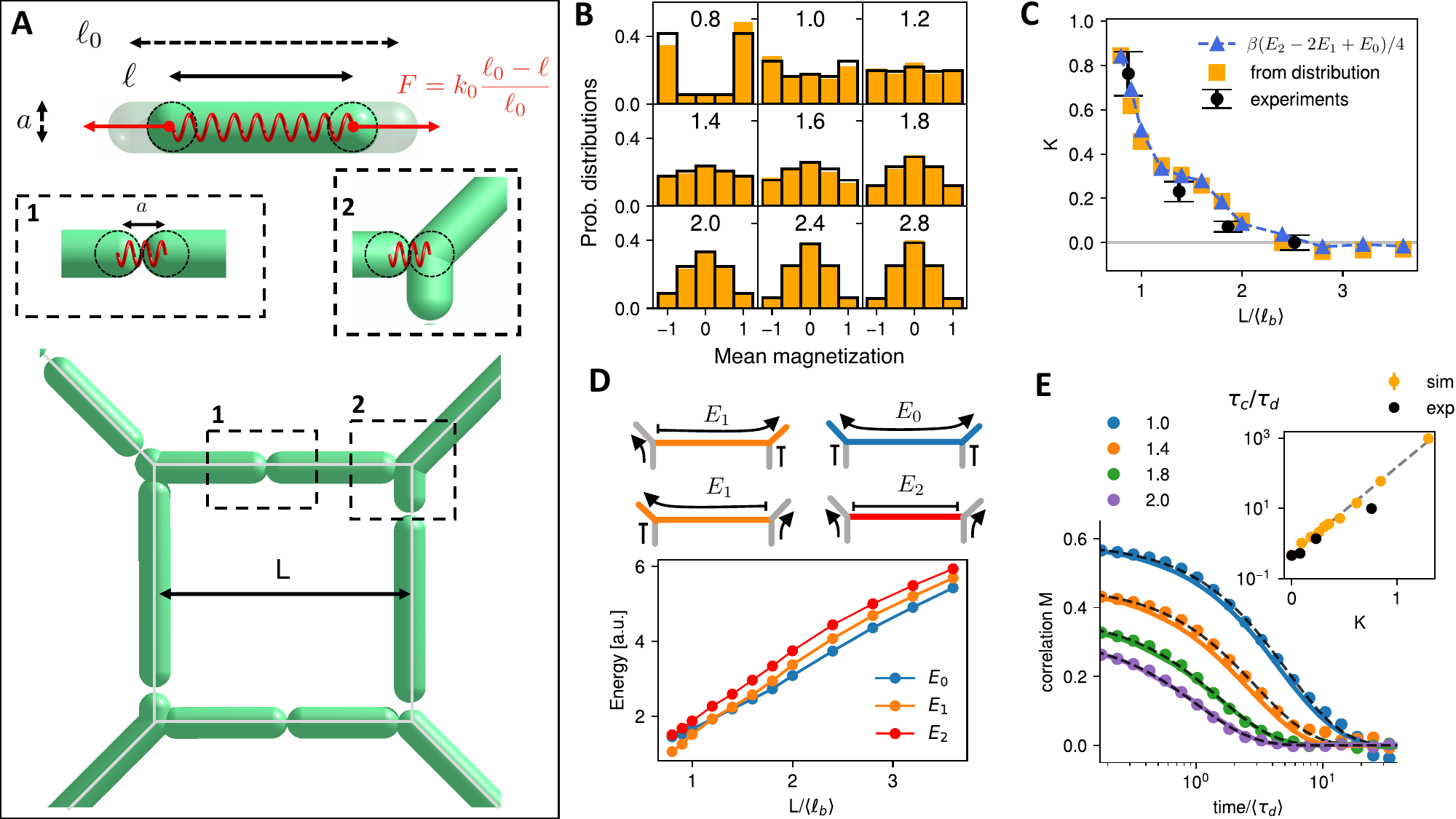}
\caption{ \textbf{Bacterial flow and spin-model analysis in simulations.} 
\textbf{(A)} Schematic of the simulation setup. Top panel: Each bacterium is modeled as two spheres of diameter \(a\) separated by a distance \(\ell\) and connected by a spring with rest length \(\ell_0\). Bottom panel: Snapshot of a simulation configuration with two zooms highlighting elastic interactions between neighboring bacteria. 
\textbf{(B)} Probability distribution of the mean magnetization for different network lengths normalized to the mean length of bacteria at birth $L/\langle \ell_b \rangle$. Orange indicates data from simulations, while the black line represents the probability distribution extracted from the model.
\textbf{(C)} Coupling parameter $K$ extracted from simulations using the average energy of the three states (blue triangles; \(\beta = 4\) in simulation units) and from the square of the mean magnetization (orange squares). Data show excellent agreement between the two methods and with experimental data when the length is normalized.
\textbf{(D)} Median energy of the three states (blocked, partially blocked, and free) as a function of normalized network size.
\textbf{(E)} Time correlation function of the mean magnetization for the four network sizes, represented by round markers. Dashed lines represent exponential fits, while solid lines show Monte Carlo dynamics extracted using the corresponding $K$ values from the statics. The inset highlights the exponential growth of the characteristic time $\tau_c$ with the parameter $K$, as obtained from simulations and experiments. Both timescales are normalized by the doubling time $\tau_d$. For panels (B,C,D,E) data points referring to simulations are averaged over more than 50 independent simulations. Standard error bars are smaller than the marker size.}
\label{fig:figsim}
\end{figure*}
%
\subsection{Minimal model validates the equilibrium description.}
We found that both statics and dynamics in our experiments are reproduced by an effective spin model, where different boundary conditions (open or closed) at the nodes of an edge result in a different ``energy" cost encoded in the spin couplings. Now, we want to investigate whether this energy cost can be attributed solely to mechanical compression work acting on elastic cell bodies. To this end, we performed numerical simulations using a minimal model of growing bacteria confined in networks, 
(see Methods and Fig.~\ref{fig:figsim}A).
Each cell is modeled as a flexible spherocylindrical body, with a relaxed total length $\ell_0$, responding elastically to normal stresses. When external confinement compresses the cell to a length $\ell$, the cell body reacts, at the two ends, with normal forces that are proportional to longitudinal strain $\epsilon=(\ell_0-\ell)/\ell_0$.   
Bacterial growth is implemented by imposing an exponential increase of the spring rest length, \(\dot{\ell}_0 = \alpha(\epsilon) \ell_0\),
with a strain dependent growth rate \(\alpha(\epsilon)\). 
When cells are unconstrained, $\epsilon=0$ and cells grow exponentially with a constant rate $\alpha_0$. As compression builds up under confinement, growth rate decreases exponentially with $\epsilon$,  
accounting for the reduction in protein production associated with cytoplasmic crowding \cite{blanc2024bacterial,alric2022macromolecular}. Each pole of a cell experiences two force contributions: an internal force generated by compression and an external force arising from repulsive interactions with neighboring cells (Fig.~\ref{fig:figsim}A).
At each time step, the position of the poles are updated using an overdamped equation of motion, such that each pole moves along the one-dimensional axes of the channels by a distance proportional to the net force acting on it.
To capture recent findings on bacterial cell-size regulation, we implemented an adder-based division rule \cite{taheri2015cell}.  Fluctuations at cell birth in both growth rate and division length were chosen to be consistent with experimental observations (Extended Data Fig.2). Additional details on the simulations are provided in the Methods.\\
Across multiple simulations, we observe a transition from uncorrelated dynamics to an ordered $M=\pm1$ state as the network size $L$ approaches the average bacterial size at birth, \(\langle \ell_b \rangle\) (Supplementary Video~2). 
As for experiments, also simulations of this non-equilibrium mechanical model confirm the validity of an effective equilibrium description. Probability distributions of $M$ are accurately reproduced by Boltzmann-like distributions (Eq. \ref{eq:probs}) with the only free parameter \(K\) constrained by Eq.~\ref{m2eq} (Fig.~\ref{fig:figsim}B). 
Unlike the abstract spin model introduced before, in which the $K$ values are input parameters with no connection to the network geometry, the mechanical model provides a prediction for the $K$ dependence on the network size $L$. When plotted as a function of the normalized network length \(L/\langle \ell_b \rangle\), simulated and measured $K$ values show an excellent agreement  (Fig.~\ref{fig:figsim}C).
This strong agreement supports the validity of the simulations and allows us to gain insight into the origin of the coupling parameter $K$.
When we first introduced the spin Hamiltonian, we defined $K\propto E_2-2E_1+E_0$ relating the coupling parameter $K$ to abstract energies $E_0, E_1, E_2$ encoding an increasing degree of internal stress associated to each edge in the network. Our working hypothesis is that we can identify these energies as the mechanical potential energy associated to internal deformation and cell-cell interactions.   
From simulation data we can assign a state 0,1,2 to each edge and compute the corresponding mechanical potential energy of its cells.
For each network size $L$, all energy values obtained across time and independent simulations were pooled, and their median was used as a representative energy (Fig.~\ref{fig:figsim}D). 
From these size dependent energies, we can define a corresponding coupling parameter $K_E(L)=\beta (E_2(L) -2 E_1(L) + E_0(L))/4$ within a multiplicative constant $\beta$.
Remarkably, \(K_E\) closely follows the trend of \(K\) inferred from the distributions when \(\beta = 4\) in simulation units (Fig.~\ref{fig:figsim}C), further supporting the validity of our equilibrium-inspired theoretical framework.
For the dynamics, we fit the magnetization correlation functions with an exponential decay, \(Ae^{-t/\tau_c}\) (dashed lines, Fig.~\ref{fig:figsim}E), and find that \(\tau_c\) increases exponentially with the coupling parameter \(K\), in agreement with experimental observations (inset). Similarly, correlation functions generated via the Monte Carlo algorithm (solid lines) closely reproduce the simulation results when the timescale is set to \(\tau_{_{\rm MC}} \approx \langle \tau_d \rangle/2\).\\
\subsection{Conclusions.} 
We studied the growth dynamics of bacterial cells competing for space in a network of narrow channels. 
We found that when edge length is comparable with cell size, the system displays collective modes of growth which can be stable over several cell cycles. As the edge lengths approach three times the cell size at birth, these correlations vanish and nodes begin to fluctuate independently.
Despite the strongly out-of-equilibrium nature of the dynamics, we were able to quantitatively capture the observed phenomenology using an effective equilibrium model in which flow configurations at each node are encoded as spin variables interacting through ferromagnetic couplings. Bacterial dynamics simulations suggest that the physical origin of the coupling can be traced back to the buildup of mechanical potential energy in cells growing within an edge with one or both nodes closed, thereby constraining the rate of outgoing biomass. For small edge lengths, correlated patterns of growth result in minimal internal stress, making them more stable than disordered patterns. Further investigations of simplified models, such as the one presented here, are essential to explore the fundamental question of the conditions under which equilibrium models can successfully provide a quantitative description of proliferating active matter. 
We expect 
that future studies will assess the general applicability of this framework in more complex network topologies, ultimately moving closer to naturally occurring scenarios.

\subsection{Acknowledgments.} The research leading to these results has received funding from the European Research Council under the ERC Grant Agreement No. 834615 (R.D.L.).



\bibliography{biblio}


\section*{Methods}
\textbf{Sample Microfabrication}
Microchambers are fabricated on soda lime glass substrates pre-cleaned for 24 hours in a solution of sulfuric acid (95–98\%) and NoChromix reagent (5\% w/v), then rinsed thoroughly in deionized water. SU-8 photoresist (~300 $\mu$l) is spin-coated first at 500 rpm for 10 s, then at 2000 rpm for 30 s, and soft-baked at 95°C for 30 min, resulting in a ~25 $\mu$m-thick layer. The microstructures are fabricated using a custom-built two-photon polymerization (TPP) setup \cite{vizsnyiczai2017light,pellicciotta2023light} with a laser scanning speed of 40 $\mu$m/s at 6 mW power. After exposure, samples undergo post-exposure baking at 95°C for 8 min, development in SU-8 developer (KAYAKU Advanced Materials) for 15 min, and nitrogen drying \cite{walther2007stability}. The glass substrate bearing the microchambers was bonded with oxygen plasma to a PDMS chip incorporating a 2 mm-wide, 200 $\mu$m-high flow channel.\\

\textbf{Cell culture}
We used \textit{E. coli} strain AB1157 deleted for \textit{cheY} to avoid bacterial tumbling, and transformed with plasmid pWR20-EGFP (a kind gift from Pietro Cicuta's Lab) to provide constitutive expression of the fluorescent reporter EGFP. More details on the strain preparation are provided in \cite{pellicciotta2025wall}. The desired strain was grown from a glycerol stock for 16 hours in LB with the appropriate antibiotics. Bacteria were refreshed 1:100 in TB under red light until they reached an OD of 0.1 (around 4 hours). 1 mL of the culture was loaded into the PDMS main channel with a pipette, and we waited for the microchamber to fill with motile bacteria (around half an hour). For the whole duration of the experiments, the chip was then perfused with LB with appropriate antibiotics and 0.1\% BSA at a constant rate of 50 $\mu$L/min.\\

\textbf{Image acquisition and data analysis}
Bright-field and epi-fluorescence imaging of GFP were performed using a custom-built optical microscope equipped with a $100\times$ magnification objective (Nikon MRH11902; NA=1.3) and a high-sensitivity CMOS camera (Hamamatsu Orca-Flash 4.0 V3) (see Supplementary Fig.~7). Additional details on the optical setup and image acquisition can be found in \cite{cannarsa2024light}.
Individual cells were segmented from the fluorescence images using the pretrained neural network model Cellpose 2.0 \cite{pachitariu2022cellpose}, which we further trained with our data to increase segmentation accuracy. The masks obtained after segmentation were used to measure the number of bacteria and the average growth rate of all cells as a function of time. For the measurement of the growth rate, length at birth, and length at division, we used a customized Python code that can track the lineage of bacteria \cite{liguori2024dynamic}. \\

\textbf{Simulations}

Each bacterium is modeled as two spheres of diameter $a$ separated by a distance $\ell$. The two spheres are connected by an elastic spring with Young's modulus $E$, cross-sectional area $\sigma$, and rest length $\ell_0$. This internal spring applies normal forces to the two spheres proportional to the longitudinal strain: $F_s = k_0 \epsilon$, where $\epsilon = (\ell_0 - \ell)/\ell_0$ is the strain and $k_0 = E/\sigma$.

Different bacteria in the network interact when the distance between the centers of two spheres (belonging to two different bacteria) is smaller than the sphere diameter, $\Delta x < a$, inducing elastic normal forces on the interacting spheres given by $F_i = k_0 (\Delta x - a)/a$. When a network node is occupied by a bacterium, the force exerted on a competing bacterium on the adjacent edge is calculated by assuming a sphere located at the node position (Fig.~\ref{fig:figsim}A).

We assume an overdamped regime, so that each sphere moves along the network edges according to $\dot{x} = \mu F$, where $\mu$ is the sphere mobility, $F$ is the total force acting on the sphere ($F_i + F_s$), and $x$ is the coordinate along the edge. For simplicity, we assume that each sphere moves only along the edges of the designed network. We additionally include Brownian noise acting on each bacterium, with diffusion coefficient $D = 0.015$ (expressed in reduced units, see below).

Bacterial growth is implemented by imposing exponential growth of the rest length over time, $\dot{\ell}_0 = \alpha(\epsilon)\ell_0$, so that in the absence of external confinement the bacterium length $\ell$ follows the exponential growth of $\ell_0$. 
We assume an exponential dependence of the growth rate on strain, $\alpha(\epsilon) = \alpha_0 e^{-\epsilon/\epsilon_0}$, with $\epsilon_0 = 1/4$, accounting for the reduction in protein production associated with cytoplasmic crowding \cite{blanc2024bacterial,alric2022macromolecular}. The intrinsic growth rate $\alpha_0$ of each bacterium is drawn from a normal distribution ($\langle \alpha_0 \rangle = 1$, $\sigma_{\alpha} = 0.3$).

Division is triggered when the spring rest length $\ell_0$ reaches a threshold corresponding to the division length $\ell_d$, 
generating two newborn bacteria with identical spring rest length $\ell_b = \ell_d/2$. For each newborn bacterium, the division length $\ell_d$ is assigned according to the adder principle \cite{taheri2015cell}, i.e., $\ell_d = \ell_b + \Delta$, where $\Delta$ is a stochastic variable drawn from a normal distribution ($\langle \Delta \rangle = 1$, $\sigma_{\Delta} = 0.3$), such that the average rest length at birth is $\langle \ell_b \rangle = 1$. The standard deviations of $\Delta$ and $\alpha_0$ are chosen according to experimental values.
All quantities are expressed in reduced units, $\langle \ell_b \rangle$ for length, $\langle \alpha_0 \rangle^{-1}$ for time and $\mu$ for mobility. In these units we set the effective spring constant $k_0 = 20$.
For each value of $L$ we performed 50 simulations of duration $T = 250$ with an integration time step $dt = 0.001$.


\end{document}